\RequirePackage[2020-02-02]{latexrelease}
\documentclass[aps,pre,graphicx,twocolumn,superscriptaddress]{revtex4-2}

\usepackage{amsmath}
\usepackage{amsfonts}
\usepackage{graphicx}
\usepackage{amssymb}
\usepackage{bm}
\usepackage{braket}
\usepackage{hyperref}
\usepackage{tikz}

 \newcommand{\YZ}[1]{\textcolor{black}{#1}}

\begin{document}

\title{Nonlinear magnetohydrodynamic modeling of current-drive-induced sawtooth-like crashes in the W7-X stellarator}

\author{Yao Zhou}
\affiliation{School of Physics and Astronomy, Institute of Natural Sciences, and MOE-LSC, Shanghai Jiao Tong University, Shanghai 200240, China}
\email[]{yao.zhou@sjtu.edu.cn}
\author{K.~Aleynikova}
\affiliation{Max-Planck-Institut f\"ur Plasmaphysik, 17491 Greifswald, Germany}
\affiliation{Max Planck/Princeton Research Center for Plasma Physics, Princeton, New Jersey 08544, USA}
\author{N.~M.~Ferraro}
\affiliation{Princeton Plasma Physics Laboratory, Princeton, New Jersey 08543, USA}

\date{\today}

\begin{abstract}
Sawtooth-like core electron temperature crashes have been observed in W7-X experiments with electron cyclotron current drive. We present nonlinear {single-fluid} magnetohydrodynamic simulations of this phenomenon using the newly developed stellarator modeling capability of the M3D-$C^1$ code. The near-axis current drive gives rise to two $\iota=1$ resonances in the equilibrium rotational transform profile so that two consecutive $(1,1)$ internal kink modes are seen in the simulations. A small-amplitude crash at the inner resonance occurs first, which may correspond to the sawtooth precursors observed in the experiments. A bigger crash at the outer resonance then flattens the core temperature profile, which shows semi-quantitative agreements with experimental measurements on {certain} metrics such as the crash amplitude and the inversion radius of the temperature change. These results illustrate {a likely} mechanism of the current-drive-induced sawtooth-like crashes in W7-X and, {to some extent,} validate the stellarator modeling capability of M3D-$C^1$.

\end{abstract}

\maketitle

\section{Introduction}
Magnetohydrodynamic (MHD) stability is essential for magnetic fusion devices including tokamaks and  stellarators.
Stellarators do not require strong plasma currents to generate the confining magnetic fields and hence are generally less susceptible to current-driven MHD instabilities than tokamaks \cite{Boozer2005}. 
That said, sometimes even small amounts of plasma currents can induce MHD events in stellarators. 
For example, the advanced W7-X stellarator is designed to minimize the plasma current, and electron cyclotron current drive (ECCD) is adopted as one of the tools for controlling the strikeline of the island divertor by compensating the bootstrap current \cite{Geiger2015}. 
However, periodic crashes of the core electron temperature have been observed in W7-X experiments with ECCD \cite{Zanini2020}. 
These crashes appear similar to the well-known sawtooth oscillations widely seen in tokamaks and are usually benign, but in limited occasions can be so strong as to prematurely terminate the plasma \cite{Zanini2021}. 
Understanding the nature of such sawtooth-like crashes in W7-X could help in avoiding them and improving confinement.

There have been extensive studies of sawtooth oscillations in tokamaks \cite{Chapman2011} {and, to a lesser extent, current-carrying stellarators \cite{Nagayama2003,Roberds2016}}. 
One leading theory is the Kadomtsev model of magnetic reconnection driven by the $(m,n)=(1,1)$ internal kink mode due to the central safety factor $q$ falling below unity \cite{Kadomtsev1976} (here $m$ and $n$ are the poloidal and toroidal mode numbers, respectively).
Such repetitive reconnection events are routinely seen in nonlinear MHD simulations at low plasma beta \cite{Jardin2012,Krebs2017,Shen2018,Zhang2020,Halpern2011} but seem elusive at high beta, where the Wesson model based on the pressure-driven interchange mode might be more applicable \cite{Wesson1986,Jardin2020}. 
The sawtooth-like crashes in W7-X occur at low beta and ray-tracing modeling predicts near-axis ECCD deposition, which produces a `hump' in the rotational transform profile with two $\iota=1$ resonances. 
Therefore, a 1D current diffusion model with Kadomtsev-type relaxations has been proposed to explain the crashes \cite{Aleynikova2021}. 
With similar $\iota$ profiles, linear stability studies have confirmed that 3D W7-X type equilibria can be kink unstable due to non-ideal (resistive, two-fluid, etc.)~effects \cite{Strumberger2020,Zocco2021,Slaby2021}, and nonlinear simulations in simplified cylindrical geometry have shown that such internal kink modes can indeed result in core temperature crashes \cite{Yu2020}. 
In contrast, nonlinear MHD simulations in W7-X geometry using rather different $\iota$ profiles with mid-radius ECCD have found that the nonlinear coupling of high-$n$ ballooning modes can induce low-$n$ modes to trigger core crashes by stochasticizing the magnetic field \cite{Suzuki2021}.

In this paper, we present nonlinear {single-fluid} MHD simulations of sawtooth-like crashes in W7-X geometry with near-axis ECCD according to ray-tracing modeling. 
The two $\iota=1$ resonances lead to two consecutive $(1,1)$ internal kink modes in the simulations. 
Reconnection at the inner resonance causes a small-amplitude crash first, which may correspond to the sawtooth precursors observed in the experiments. 
A bigger crash at the outer resonance then flattens the core temperature profile, which shows semi-quantitative agreements with experimental measurements on {certain} metrics such as the crash amplitude and the inversion radius of the temperature change. 
These results suggest that the mechanism of the sawtooth-like crashes in W7-X is likely {reconnection driven by $(1,1)$ internal kink modes.} 
{To some extent}, they also validate the newly developed stellarator modeling capability of the M3D-$C^1$ code \cite{Zhou2021}, which enables nonlinear MHD modeling of {complex} stellarators at transport timescales.

This paper is organized as follows. In Section \ref{model} we describe the numerical model used in the simulations. In Section \ref{equilibrium} we prepare the equilibrium we initialize the simulations with. In Section \ref{results} we present the simulation results and compare with experimental measurements. Summary and discussion follow in Section \ref{summary}.

\section{Numerical model}\label{model}
M3D-$C^1$ is a sophisticated nonlinear MHD code that has mainly been used to model the macroscopic dynamics of tokamak plasmas \cite{Jardin2012}. 
For time advance, M3D-$C^1$ implements a split-implicit scheme that allows for time steps larger than Alfv\'enic \cite{Jardin2012b}, which realizes stable and efficient transport-timescale simulations. 
For spatial discretization, M3D-$C^1$ uses high-order finite elements with $C^1$ continuity in all three dimensions and has recently been extended to treat non-axisymmetric stellarator geometry \cite{Zhou2021}. 
While two-fluid and many other effects are available in M3D-$C^1$, they are not yet functional in  stellarator geometry. 
So in this work, we solve the single-fluid extended MHD equations, including the momentum equation for the fluid velocity $\mathbf{v}$ (in SI units)
\begin{gather}\label{momentum}
\rho(\partial_t \mathbf{v} + \mathbf{v}\cdot\nabla\mathbf{v}) = \mathbf{j}\times\mathbf{B} - \nabla p - \nabla\cdot\mathbf{\Pi},
\end{gather}
the energy equation for the fluid pressure $p$
\begin{align}\label{energy}
\partial_t p + \mathbf{v}\cdot\nabla p +\Gamma p\nabla\cdot\mathbf{v} &= \nonumber\\
(\Gamma-1)(\eta j^2 &- \nabla\cdot\mathbf{q} - \Pi : \nabla\mathbf{v}+Q),
\end{align}
and the induction equation for the magnetic field $\mathbf{B}$
\begin{gather}\label{induction}
\partial_t \mathbf{B} =  \nabla\times[\mathbf{v}\times\mathbf{B} - \eta (\mathbf{j}-\mathbf{j}_0)],
\end{gather}
where the current density $ \mathbf{j}$ is given by Amp\`ere's law, $\mu_0 \mathbf{j}= \nabla\times\mathbf{B}  $. 
The stress tensor is given by $\mathbf{\Pi} = -\mu(\nabla\mathbf{v}+\nabla\mathbf{v}^{\text{T}}) - 2(\mu_{\text{c}}-\mu)(\nabla\cdot\mathbf{v})\mathbf{I}$ and the heat flux $\mathbf{q} = -\kappa_\perp\nabla T - \kappa_\parallel\mathbf{b}\mathbf{b}\cdot\nabla T$, with $\mathbf{b} =\mathbf{B}/B$ and the temperature $T=Mp/\rho$, where $M$ is the ion mass. 
Transport coefficients include resistivity $\eta$, isotropic and compressible viscosities $\mu$ and $\mu_{\text{c}}$, and perpendicular and parallel thermal conductivities $\kappa_\perp$ and $\kappa_\parallel$, and $\Gamma=5/3$ is the adiabatic index. 
In addition, $Q$ is the heat source and the ECCD current density $\mathbf{j}_0=j_\text{CD}\bm{\varphi}$ effectively acts as a current source in the toroidal direction $\bm{\varphi}$ in cylindrical coordinates $(R,\varphi,Z)$. 
{According to \cite{Zanini2020}, the bootstrap current is relatively small in the experiments and yields a negligible contribution in the crash region}, and hence not considered in the simulations.

Note that here we do not solve the continuity equation but hold the mass density $\rho$ constant such that \eqref{energy} is essentially a temperature equation. 
This is a reasonable approximation since the core density profile stays flat and relatively unchanged in the experiments that we model, while we find such a profile difficult to maintain when we actually solve the continuity equation. 

\section{Equilibrium preparation}\label{equilibrium}
It is convenient to initialize a fixed-boundary stellarator simulation in M3D-$C^1$ using the output of the widely used 3D equilibrium code VMEC \cite{Hirshman1983}, including the geometry of the flux surfaces as well as the magnetic field and the pressure.
The former is given in terms of a coordinate mapping, $R(s,\theta,\varphi)$ and $Z(s,\theta,\varphi)$ with $s$ being the normalized toroidal flux and $\theta$ the poloidal angle in VMEC, which is utilized to set up the non-axisymmetric computational domain {enclosed by the last closed flux surface}. (This mapping does not evolve in M3D-$C^1$ so that $s$ and $\theta$ are fixed regardless of the actual dynamics of the flux surfaces.)
The latter then provide the initial conditions in the M3D-$C^1$ simulation. 
{The boundary conditions are ideal on the magnetic field, no-slip on the velocity, and fixed on the pressure.}
Since VMEC assumes nested flux surfaces based on ideal MHD while M3D-$C^1$ includes dissipation and sources, for self-consistency, it is important to ensure that the VMEC equilibrium can be approximately sustained in M3D-$C^1$ at transport timescale. 
The procedure to prepare such an equilibrium is as follows. 

\begin{figure}
\includegraphics[scale=0.45]{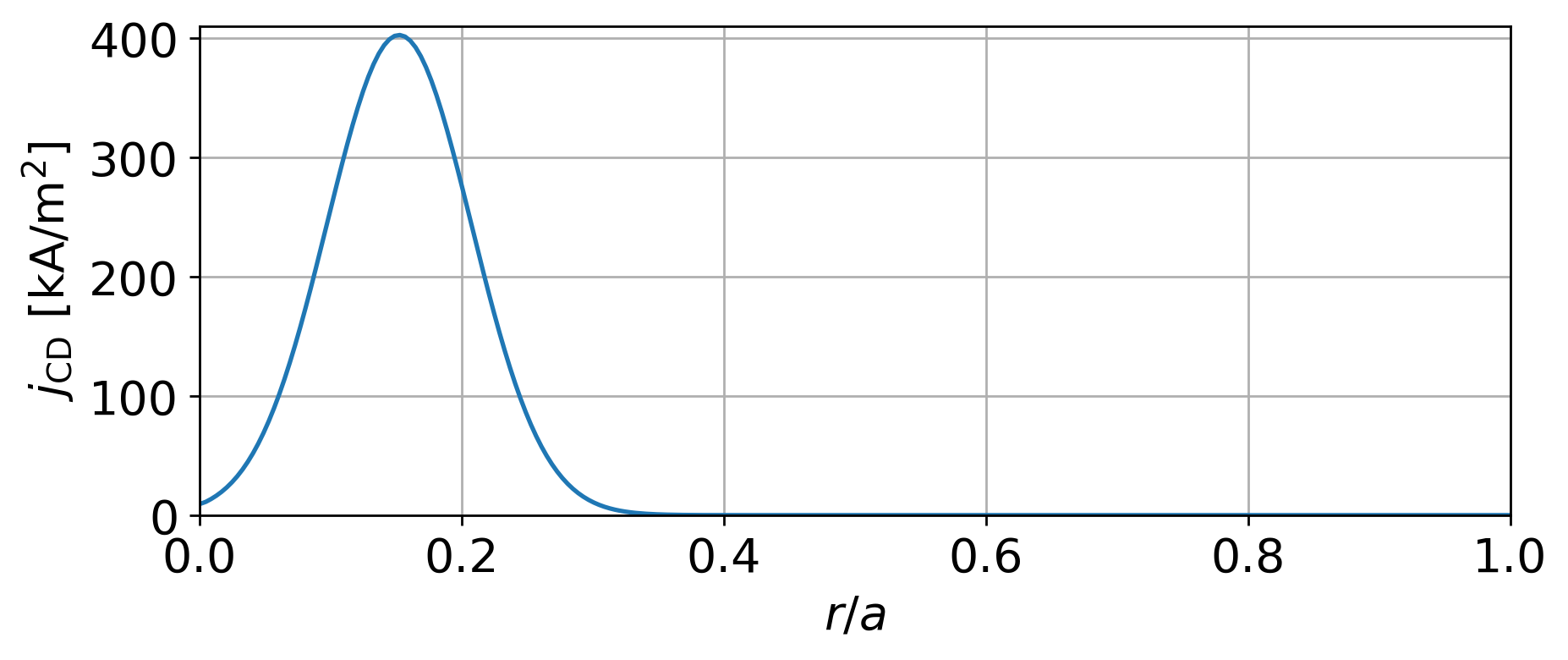}
\caption{\label{cd_profile}The ECCD profile obtained from ray-tracing modeling and used in our simulations, which peaks at $r/a\approx 0.15$. }
\end{figure}

First, we construct a W7-X type VMEC equilibrium with the best guesses for the rotational transform and pressure profiles. 
The pressure profile is extracted directly from experimental data, but there is no measurement of rotational transform available yet on W7-X. 
Instead, a rotational transform profile can be obtained by solving a reduced 1D current diffusion equation as in \cite{Aleynikova2021} using the ECCD profile $j_\text{CD}$ given by the ray-tracing code TRAVIS \cite{Marushchenko2014}, which is shown in Figure \ref{cd_profile}. {Note that we plot all profiles with respect to the normalized minor radius $r/a=\sqrt{s}$, which serves as a proxy for comparison with experimental results. For reference, the averaged major and minor radii in the VMEC equilibrium are 5.51 and 0.506 m, respectively, and the total toroidal flux enclosed is 2.133 Wb.}

Next, we use this VMEC equilibrium to initialize M3D-$C^1$ simulations in a single field period, i.e., one fifth of the full torus in W7-X. The simulations are run for sufficient time to reach saturated states. Even though we use the same $j_\text{CD}$ as in Figure \ref{cd_profile}, the rotational transform profile would still evolve slightly since the M3D-$C^1$ model is different from the 1D current diffusion equation. Meanwhile, we need to find by trial and error a combination of heat source $Q$ and thermal conductivities $\kappa_\perp$ and $\kappa_\parallel$ that can roughly maintain the initial pressure profile.

\begin{figure}
\includegraphics[scale=0.45]{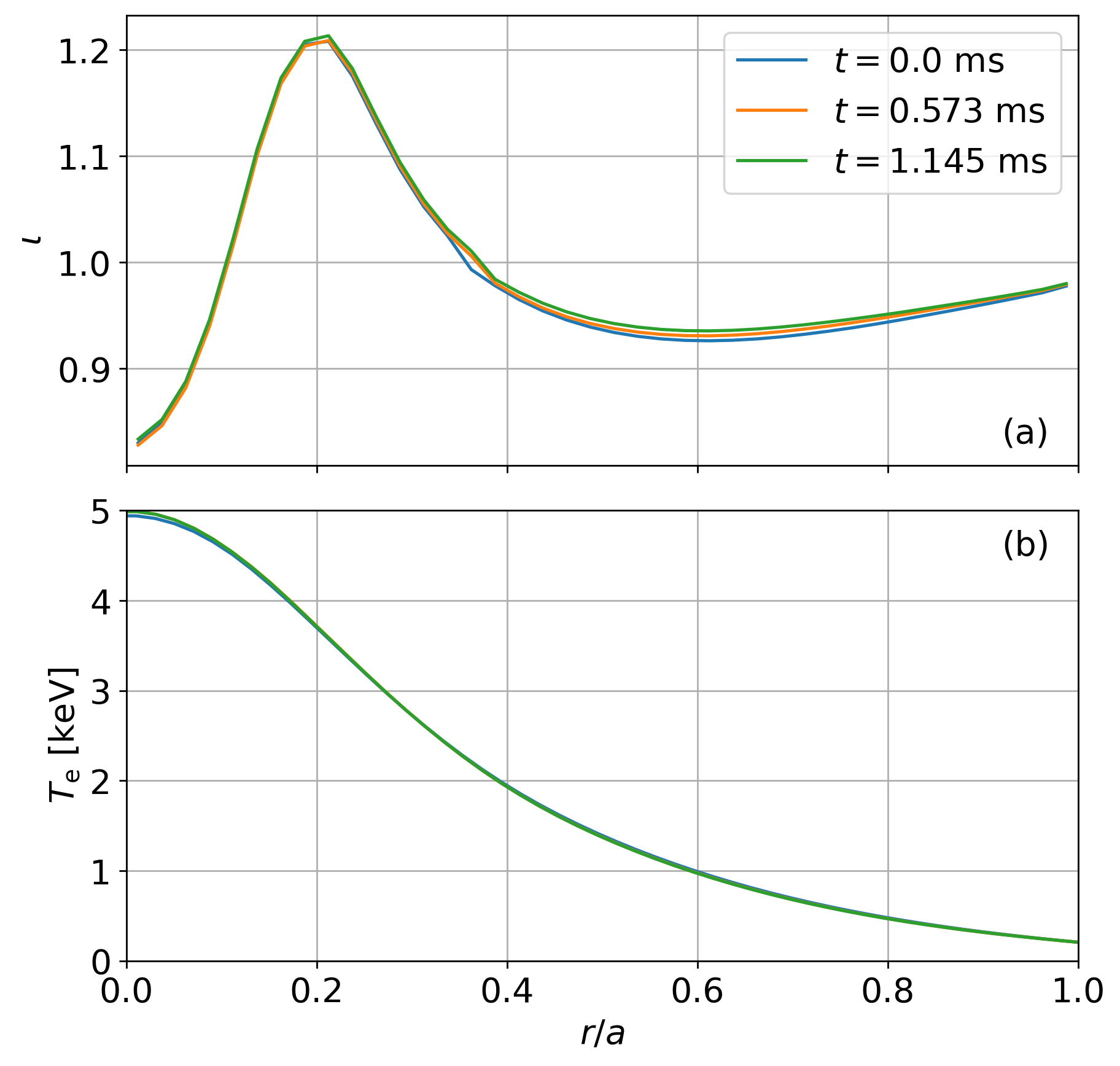}
\includegraphics[scale=0.45]{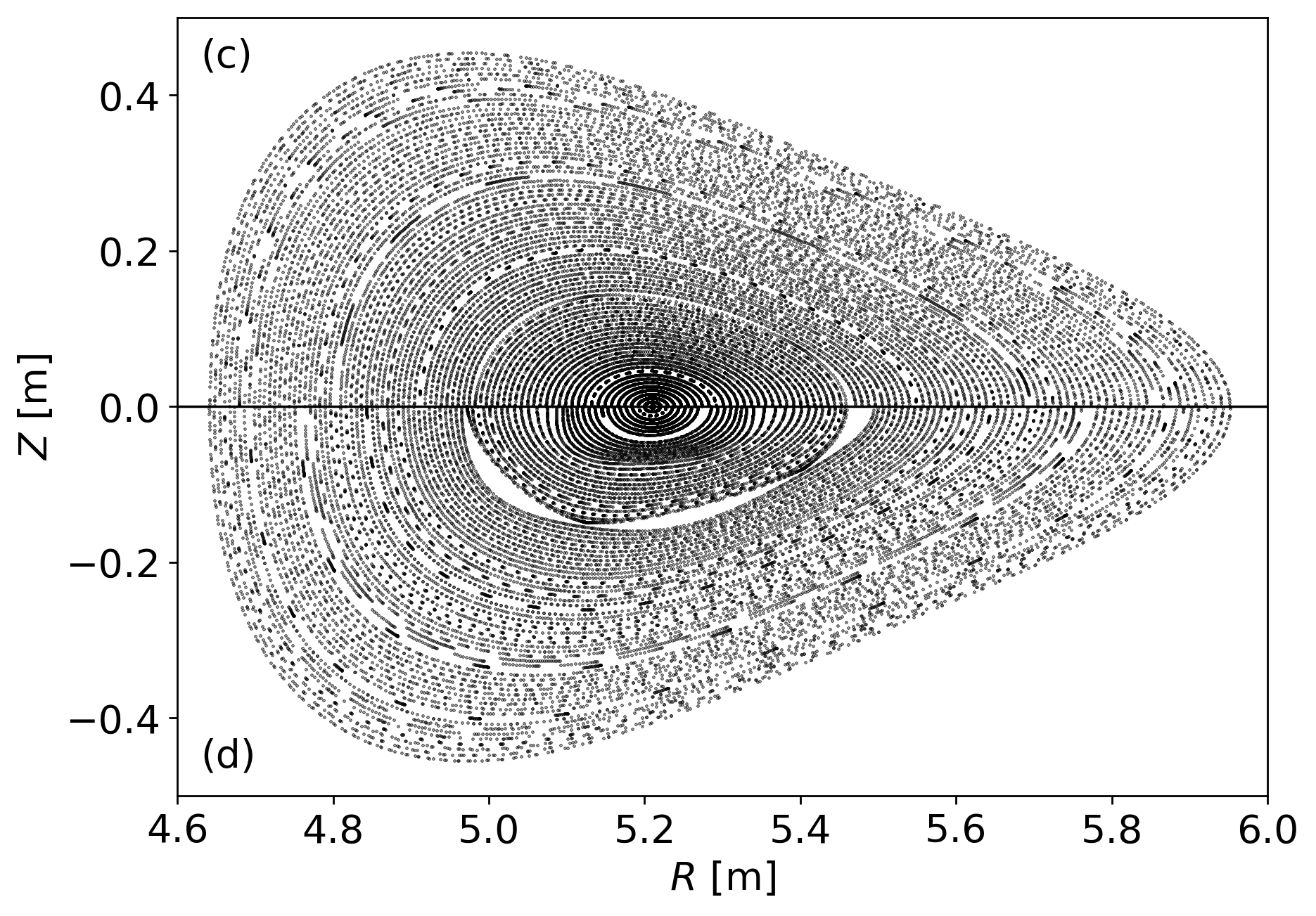}
\caption{\label{single}Results from the final one-field-period M3D-$C^1$ simulation. Snapshots of (a) the rotational transform obtained from field-line tracing and (b) the electron temperature at $\theta=0$ and $\varphi=0$ show that the profiles barely evolve. The Poincar\'e plots at $\varphi=\pi$ and (c) $t=0$ ms and (d) $t=1.145$ ms show limited change in field-line topology.}
\end{figure}

Finally, we use the rotational transform and pressure profiles from the saturated state to create a new VMEC equilibrium, with which we initialize another one-field-period M3D-$C^1$ simulation to check whether it can be consistently sustained with the same sources and transport coefficients. 
Indeed, this is demonstrated by the results presented in Figure \ref{single}. 
It can be seen that the rotational transform and electron temperature $T_\text{e}$ profiles barely evolve, and that the change in field-line topology is limited to the formation of some small island chains, in particular an $(m,n)=(5,5)$ one at the outer $\iota=1$ resonance.

The parameters used in this simulation are specified as follows. 
We use uniform viscosities $\mu=\mu_\text{c}=3.65\times10^{-5}$ $\mathrm{kg/(m\cdot s)}$ and resistivity $\eta=2.74\times10^{-6}~\mathrm{\Omega\cdot m}$, which are enhanced from realistic values to make the simulation stable and practical. The parallel thermal conductivity $\kappa_{\parallel}=2.18\times10^{26}$ $\mathrm{(m\cdot s)^{-1}}$, while the perpendicular thermal conductivity $\kappa_{\perp}$ depends linearly on $T_\text{e}^{-1/2}$ and its equilibrium value varies from $4.03\times10^{19}$ $\mathrm{(m\cdot s)^{-1}}$ at the center to $1.12\times10^{20}$ $\mathrm{(m\cdot s)^{-1}}$ at the boundary. 
The heat source is given by $Q=w/(2\pi\sigma^2)\,\mathrm{e}^{-s/(2\sigma^2)}$ with strength $w=9.73\times10^{6}$ $\mathrm{Pa/s}$ and width $\sigma=0.15$.
To mimic the experiments, we consider a hydrogen plasma with a number density of $2\times10^{19}$ $\mathrm{m}^{-3}$ and set the energy partition $T_\text{e}/T=5/6$. 
We have tried varying the parameters and found qualitatively similar results, and what is presented here is the most representative.
We use 3807 reduced quintic elements in the $(R,Z)$ plane and 16 Hermite cubic elements in the toroidal direction, and the size of the time step is $0.573~\mu s$. Note that the relatively high toroidal resolution (for a single field period) is needed to accurately treat the strongly anisotropic heat transport.

\section{Simulation results}\label{results}

Now, to simulate a sawtooth-like crash in W7-X, we initialize a full-torus M3D-$C^1$ simulation from the final state of the one-field-period simulation shown in Figure \ref{single} ($t=1.145$ ms). 
All the simulation settings are kept the same except that the toroidal resolution is increased by five-fold (80 Hermite cubic elements) and that a small $(m,n)=(1,1)$ perturbation is applied to the velocity to speed up the onset of the crash. Simulations without the $(1,1)$ velocity perturbation (not shown) produce essentially the same crash features, including crash time and crash amplitude, etc.

\begin{figure}
\includegraphics[scale=0.48]{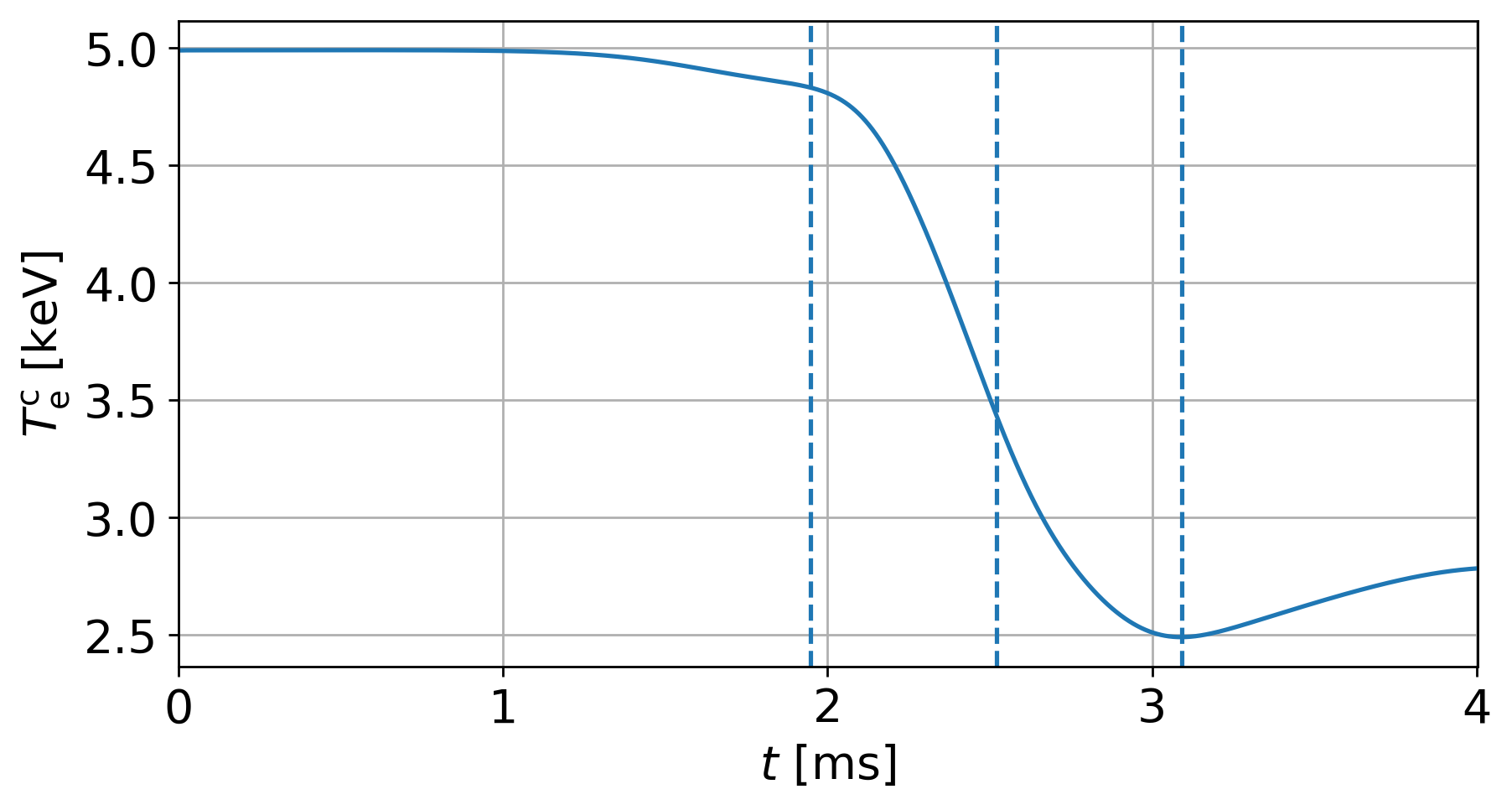}
\caption{\label{temax}The core electron temperature at a fixed \YZ{location} ($s=0,~\varphi=0$) versus time in the full-torus M3D-$C^1$ simulation. Two crashes can be seen: a smaller one beginning at $t\approx1.2$ ms, and a bigger one beginning at $t\approx2$ ms. The vertical dashed lines mark the instants shown in Figures \ref{outer} and \ref{crash}.}
\end{figure}

\begin{figure}
\includegraphics[scale=0.45]{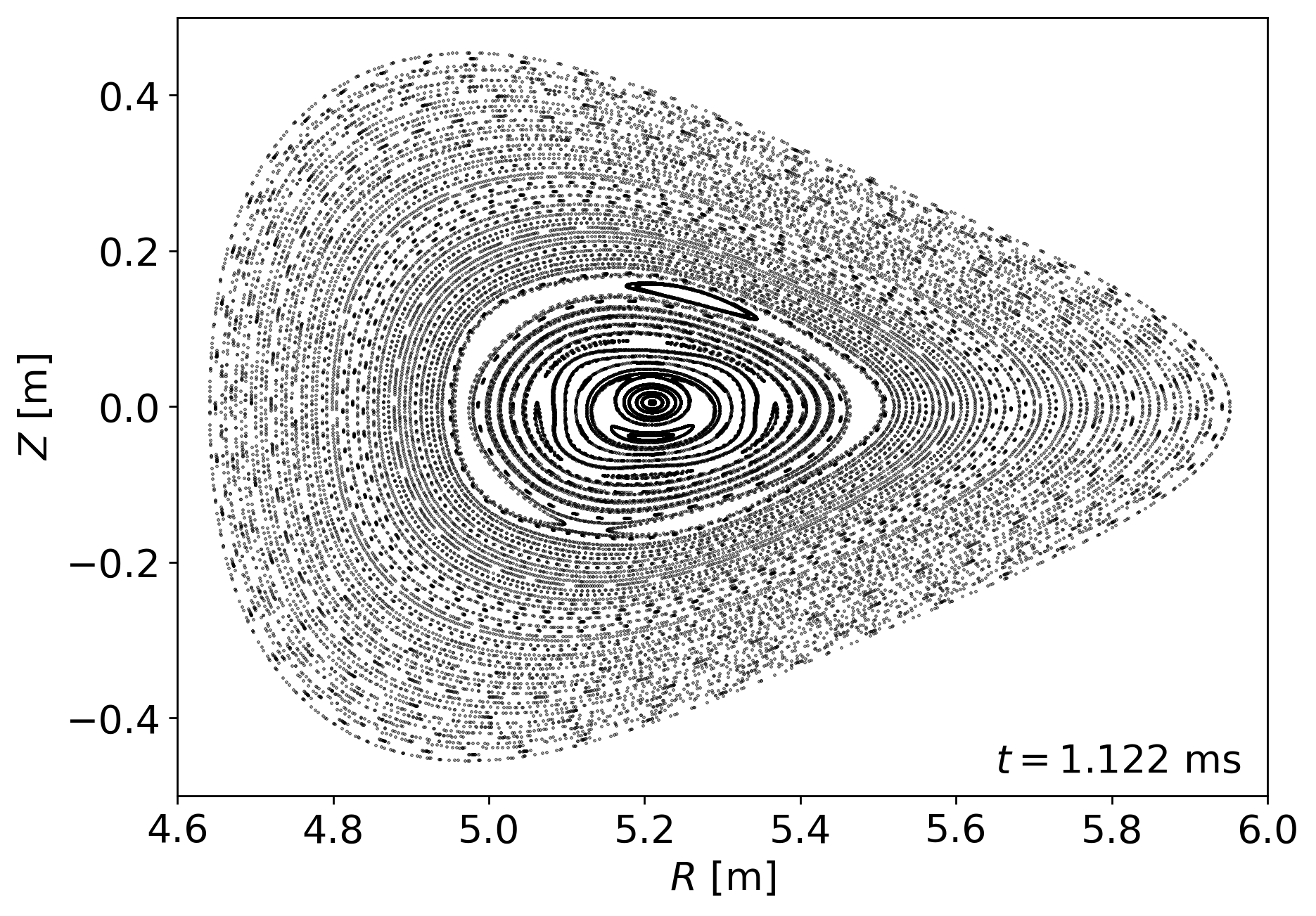}
\includegraphics[scale=0.45]{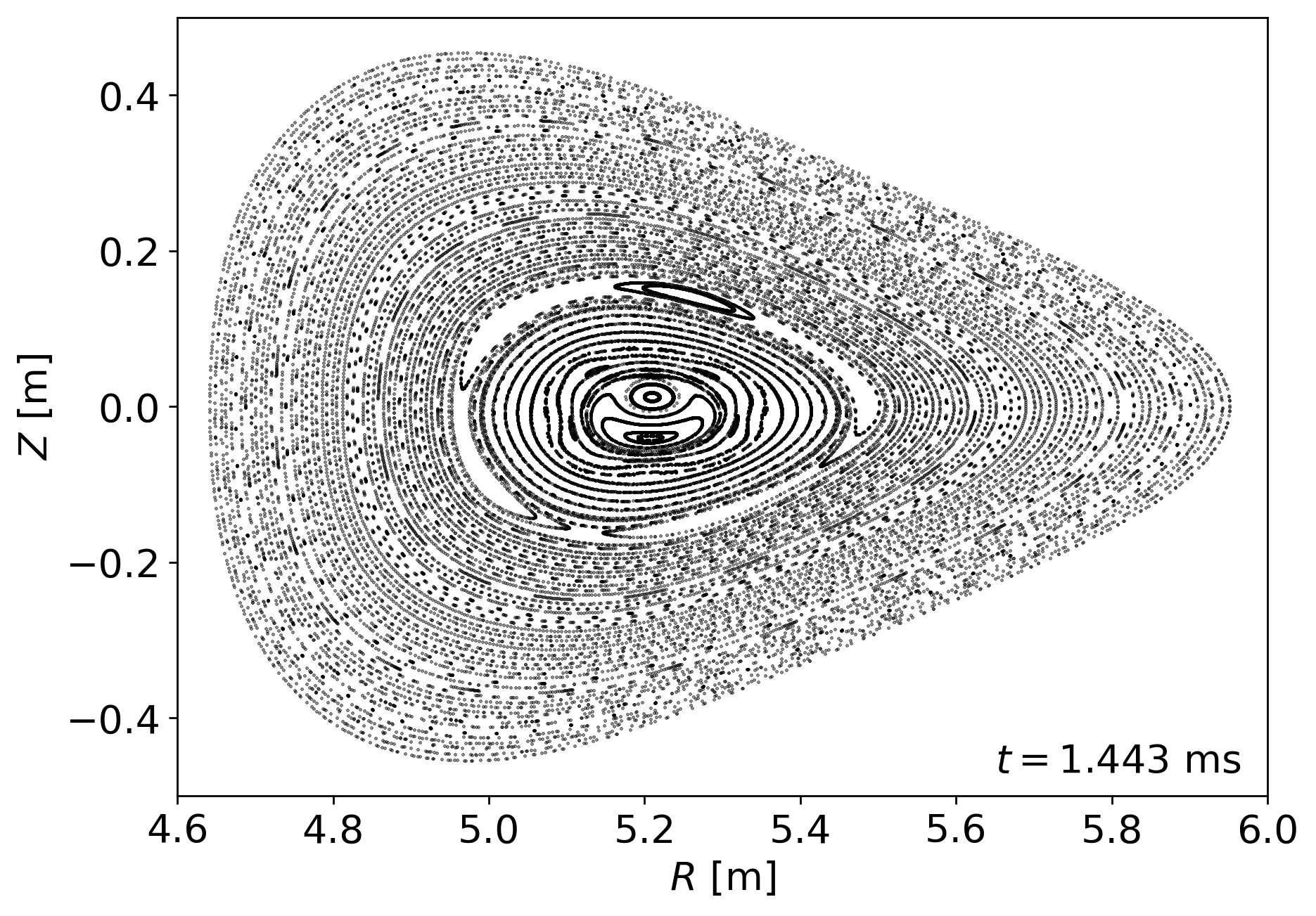}
\includegraphics[scale=0.45]{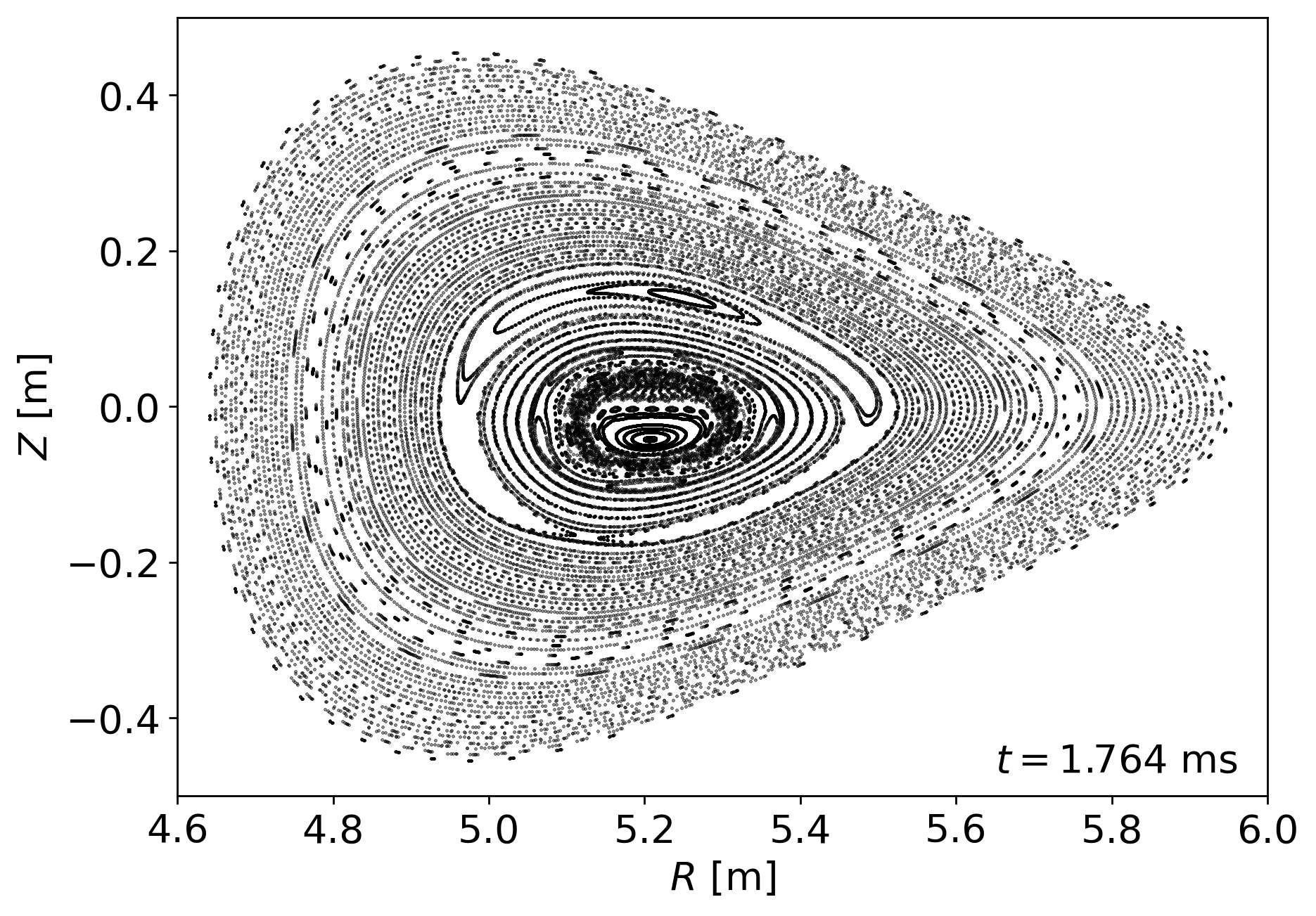}
\caption{\label{inner}Snapshots of Poincar\'e plots at $\varphi=\pi$ during the inner crash show  `upward' core displacement and magnetic reconnection at the inner $\iota=1$ resonance.}
\end{figure}

The time trace of the core electron temperature (at a fixed \YZ{location}) $T_\text{e}^\text{c}$ is shown in Figure \ref{temax}. At first, $T_\text{e}^\text{c}$ remains almost constant until $t\approx1.2$ ms. 
Then, from $t\approx1.2$ ms to $t\approx2$ ms, $T_\text{e}^\text{c}$ decreases slightly from 5 keV to about 4.8 keV. 
We shall refer to this phase as the ``inner" crash because of the evolution of the magnetic field configuration shown in Figure \ref{inner} {(and also the supplementary video)}.  
Clearly, this crash is due to the growth of a $(1,1)$ island at the inner $\iota=1$ resonance, which expels and eventually overtakes the original core. 
These features are signatures of Kadomtsev's sawtooth model based on magnetic reconnection driven by the $(1,1)$ internal kink mode. 
However, this mode only impacts a limited region of $r/a\lesssim 0.1$ and hence only causes a small core temperature drop.
We speculate that the inner crash may correspond to the sawtooth precursors sometimes observed to precede the ``type-A''  crashes \cite{Zanini2020}.

\begin{figure}
\includegraphics[scale=0.45]{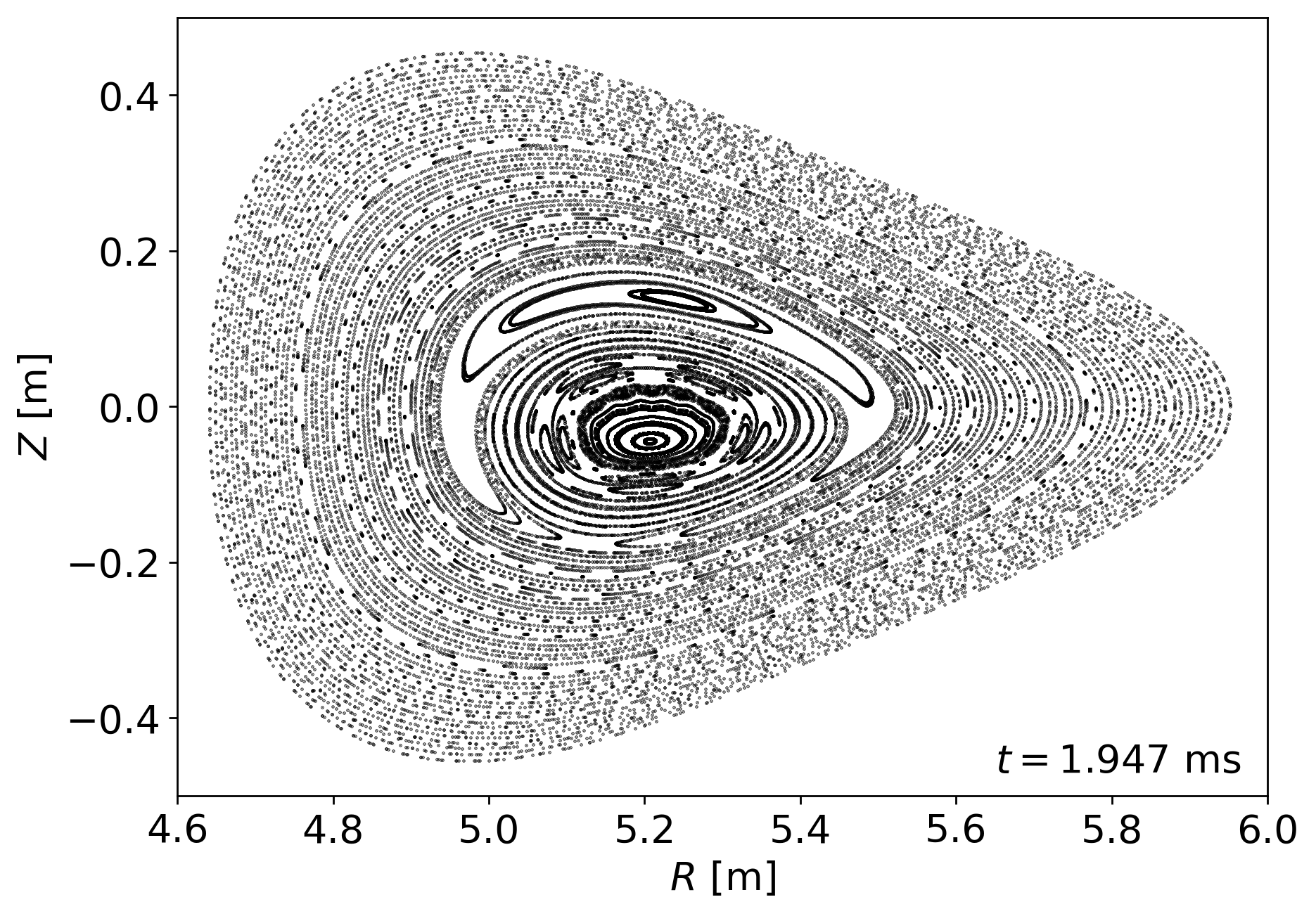}
\includegraphics[scale=0.45]{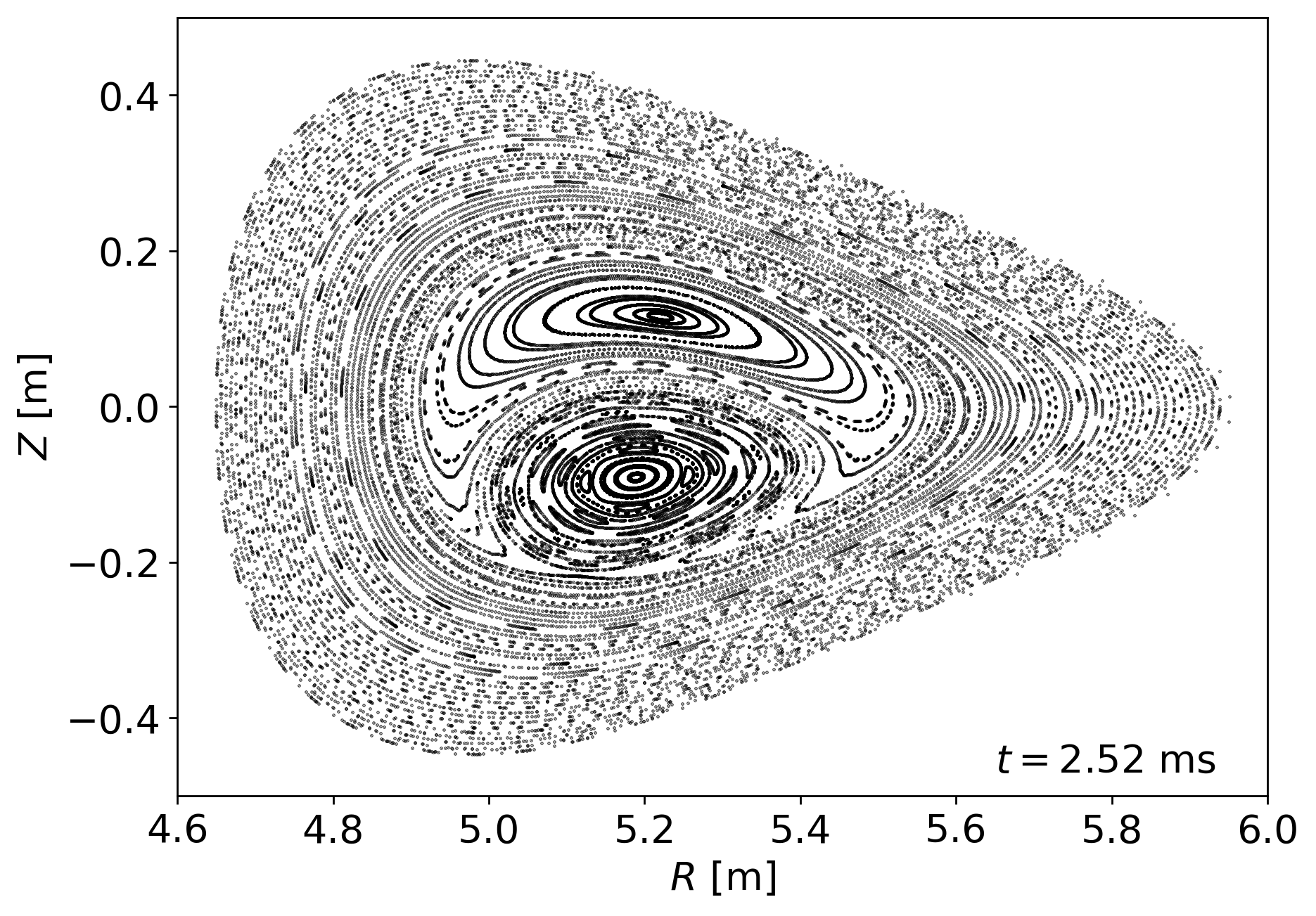}
\includegraphics[scale=0.45]{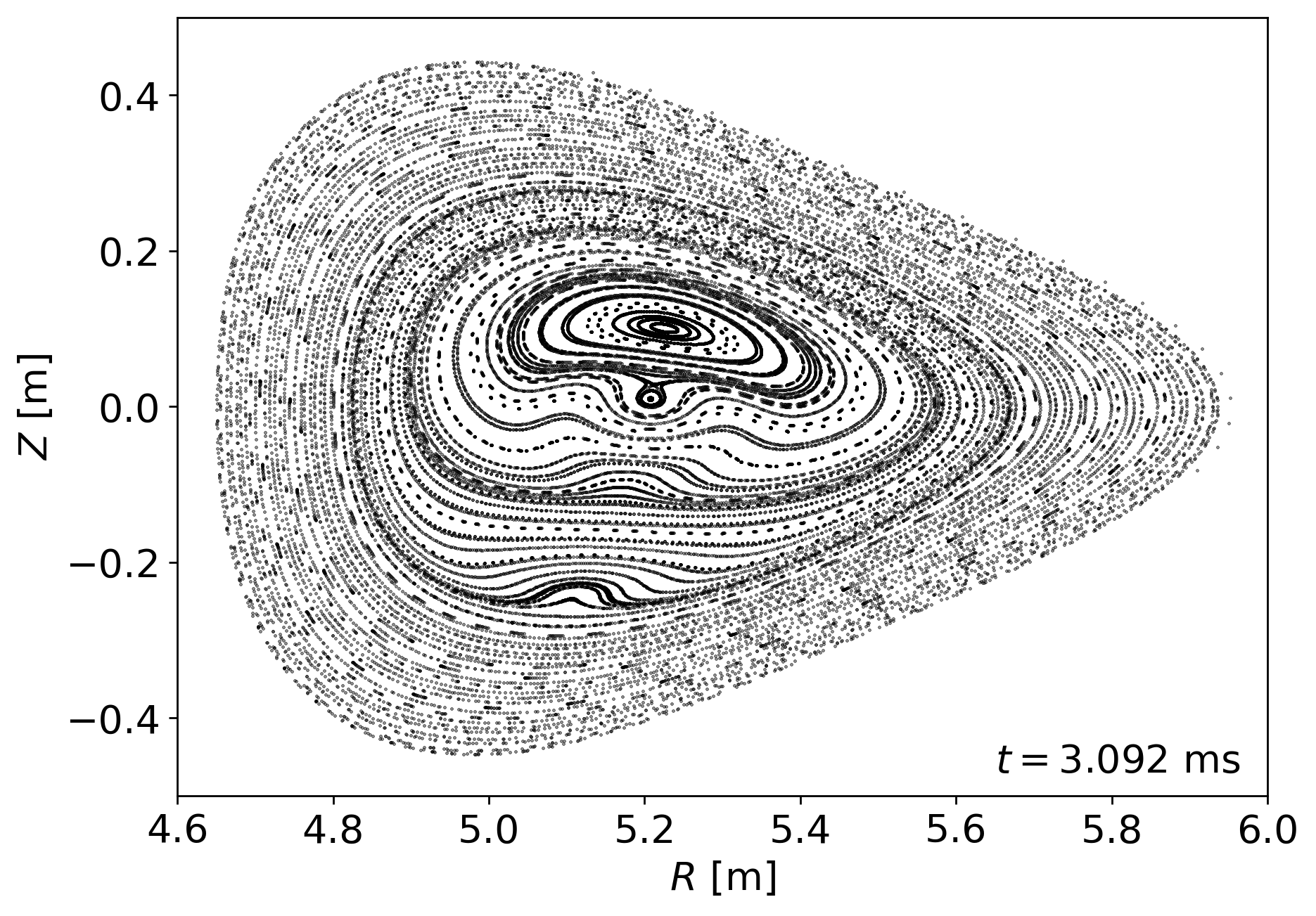}
\caption{\label{outer}Snapshots of Poincar\'e plots at $\varphi=\pi$ during the outer crash show `downward' core displacement and magnetic reconnection at the outer $\iota=1$ resonance.}
\end{figure}

Subsequently, from $t\approx2$ ms to $t\approx3$ ms, $T_\text{e}^\text{c}$ drops significantly to less than 2.5 keV, after which $T_\text{e}^\text{c}$ starts to slowly increase.
The evolution of the magnetic field configuration during this phase, which we refer to as the ``outer'' crash, is shown in Figure \ref{outer} {(and also the supplementary video)}. 
We can see that the newly formed core moves in the downward direction (in this particular plane), which is opposite to the upward displacement of the original core, and is eventually eliminated as a $(1,1)$ island at the outer $\iota=1$ resonance grows significantly. 
This internal kink mode displaces a much bigger fraction of the plasma and hence is able to cause a substantial core temperature crash. 
These features suggest that the ``type-A'' sawtooth-like crashes in W7-X are likely {due to reconnection driven by $(1,1)$ internal kink modes.}

\begin{figure}
\includegraphics[scale=0.48]{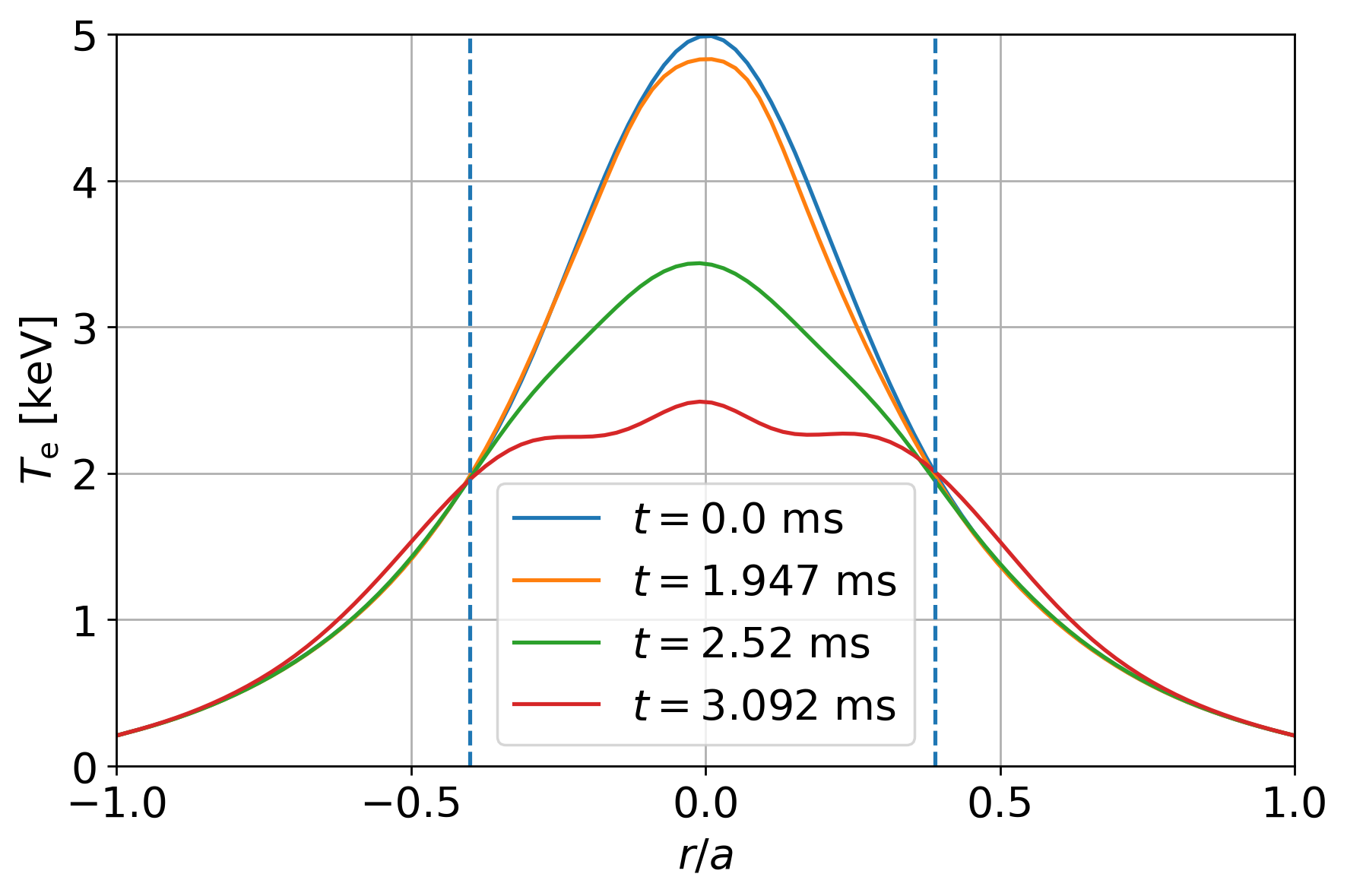}
\caption{\label{crash}Snapshots of the electron temperature profile at $Z=0$ and $\varphi=0.1$ before, during, and after the outer crash. The equilibrium profile is also shown for comparison. The vertical dashed lines mark the inversion radius of the temperature change, $r_\text{inv}/a\approx 0.4$.}
\end{figure}

The evolution of the temperature profile is shown in Figure \ref{crash} and can be compared against {} experimental results. 
The amplitude of the outer temperature crash is similar to typical values of the medium, ``type-A'' crashes in the experiments, which is $\gtrsim 50\%$ \cite{Zanini2020} (these crashes are `medium' by contrast to the major, shot-terminating crashes discussed in \cite{Zanini2021}). 
Another key metric here is the inversion radius $r_\text{inv}$ of the temperature change, which is positive at $r>r_\text{inv}$ and negative at $r<r_\text{inv}$. It can be seen that $r_\text{inv}/a\approx 0.4$ in this simulation, which is also consistent with typical experimental measurements \cite{Zanini2020}. {To some extent}, these semi-quantitative agreements validate the stellarator modeling capability of M3D-$C^1$.

\begin{figure}
\includegraphics[scale=0.45]{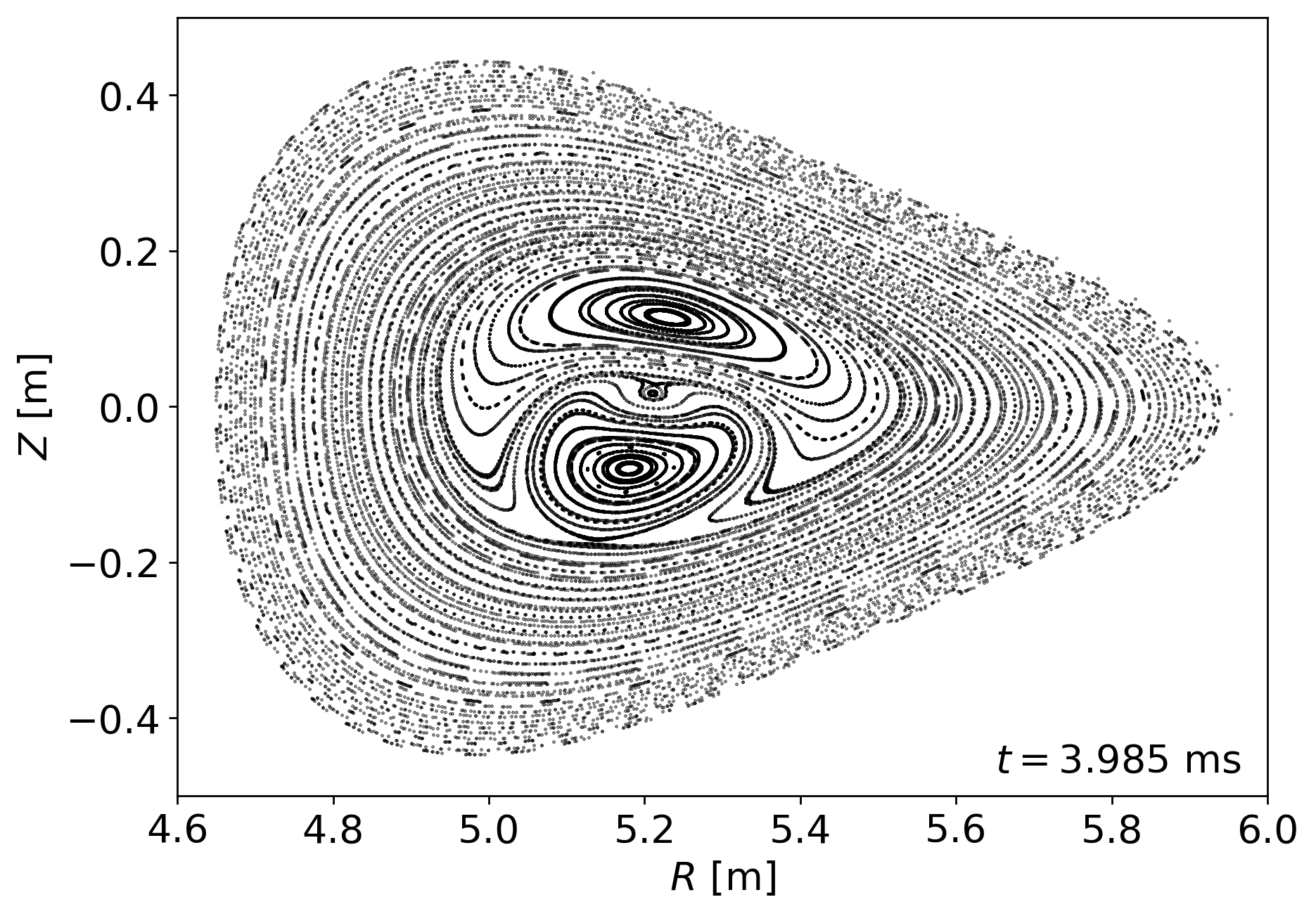}
\caption{\label{helical}A snapshot of the Poincar\'e plot at $\varphi=\pi$ in the reheating phase shows a saturated (1,1) structure in the core.}
\end{figure}

There are aspects in which the simulation does not agree quite well with the experiments, most notably the timescale of the crash. 
In the simulation the outer crash takes $\gtrsim 1$ ms whereas typical experimental values are $\lesssim 100~\mu\text{s}$. 
This is not surprising since it is well documented that purely resistive reconnection cannot account for the fast sawtooth crash \cite{Jardin2012,Krebs2017,Shen2018,Zhang2020}, and here the relatively large viscosity likely slows it down further. 
To reproduce the experimentally measured timescale, some sort of fast-reconnection mechanism may be needed, such as two-fluid effects \cite{Beidler2017} or the plasmoid instability \cite{Gunter2015}.

\begin{figure}
\includegraphics[scale=0.48]{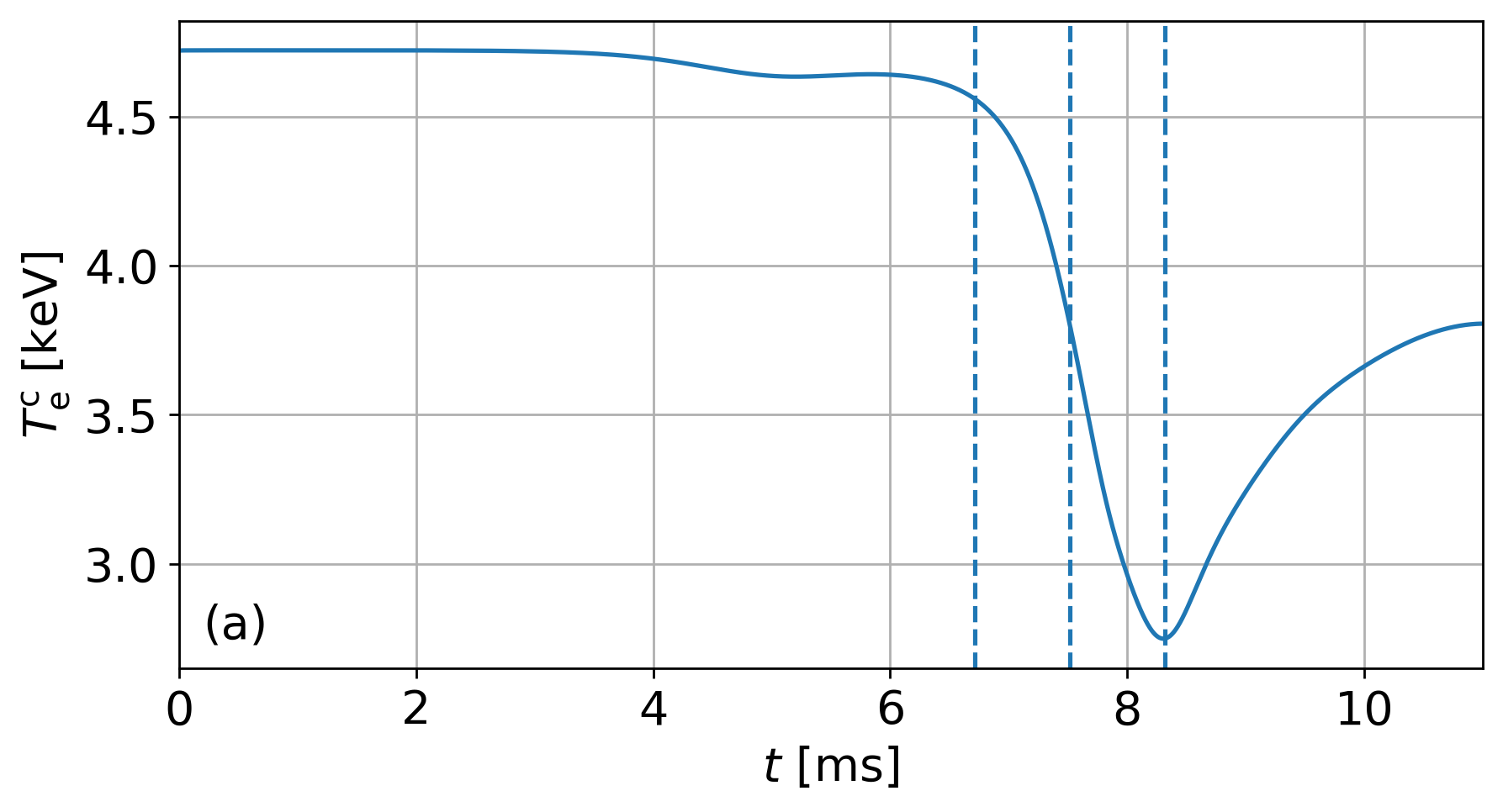}
\includegraphics[scale=0.48]{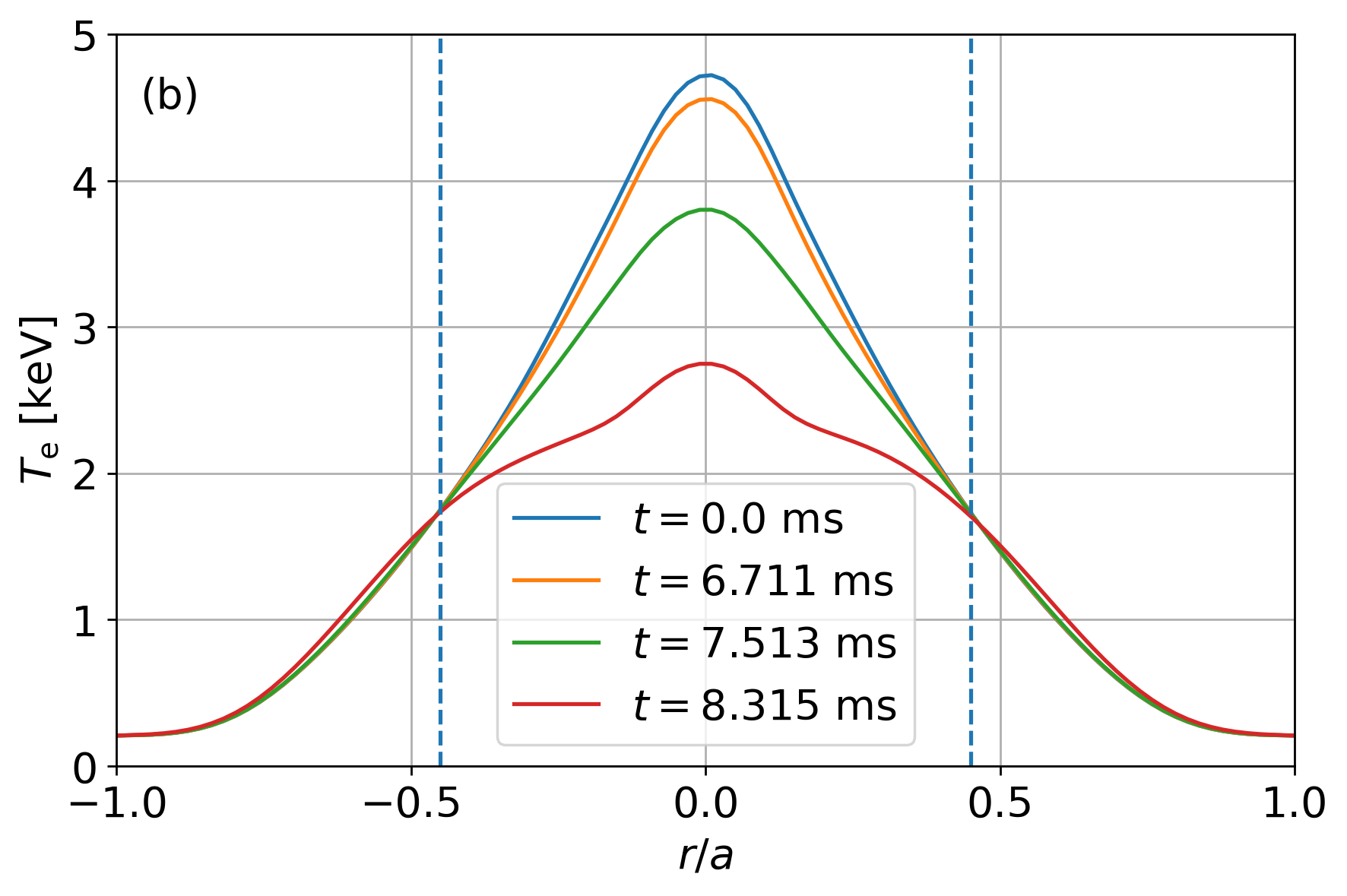}
\caption{\label{crash2}Results from a different full-torus simulation with a temperature-dependent resistivity: (a) core electron temperature versus time, similar to Figure \ref{temax}; (b) snapshots of the electron temperature profile before, during, and after the outer crash, similar to Figure \ref{crash}. The vertical dashed lines in (a) mark the instants shown in (b). The vertical dashed lines in (b) mark the inversion radius, $r_\text{inv}/a\approx 0.45$.}
\end{figure}

Another subtlety is that after the crash, the plasma does not tend to restore the initial equilibrium as one would expect in a sawtooth cycle. 
Instead, a saturated (1,1) structure forms in the core as shown in Figure \ref{helical}. 
This phenomenon is fairly common in tokamak simulations and related to the so-called ``magnetic flux pumping'' mechanism \cite{Krebs2017}. 
The formation or not of such structures can be sensitive to the transport parameters \cite{Shen2018,Zhang2020,Halpern2011}, and of particular importance here is the resistivity, to which the strength of the ECCD is proportional. 
However, we consistently obtain similar saturated (1,1) structures in simulations with somewhat varied settings.
An example with closer-to-realistic, temperature-dependent resistivity is presented in Figure \ref{crash2}. 
Specifically, $\eta$ depends linearly on $T_\text{e}^{-3/2}$ and the equilibrium value varies from $7\times10^{-7}~\mathrm{\Omega\cdot m}$ at the center to $7.5\times10^{-5}~\mathrm{\Omega\cdot m}$ at the boundary. In addition, the current and heat sources are slightly different such that the equilibrium $\iota$ profile has a smaller hump ($\iota_\text{max}\approx1.18$) and temperature profile is more peaked.
Here, the smaller core resistivity further slows down the crash. 
Hence, the crash amplitude is reduced, yet the inversion radius $r_\text{inv}/a\approx 0.45$ is still reasonably close to experimental values. 
Nonetheless, we do find that the (1,1) structure does not form if the ECCD is turned off during the crash (not shown). This raises the possibility that a more accurate ECCD model, for example one evolving with the  dynamical magnetic field rather than fixed in space, could help in reproducing the cyclic behavior.

A related issue is the accessibility of the unstable equilibrium we initialize the simulation with. 
While we obtain the equilibrium using one-field-period simulations, in a full torus instabilities could kick in before the equilibrium is reached and redistribute the current and pressure. 
In fact, there are small, ``type-B'' crashes that frequently occur between the medium, ``type-A'' crashes in the experiments \cite{Zanini2020}. In \cite{Aleynikova2021} the authors suggest that the small crashes are associated with another resonance, $\iota=5/6$. 
The results of their model, based on simulating resistive current diffusion including Kadomtsev-like crashes, show reasonable agreement with the experimental observations. 
 {An $\iota=5/6$ resonance very close to the axis is present in our simulations and has no noticeable effect on the results, but this does not mean that it cannot play a role in the build-up process of the equilibrium, for example, from a current-free state.
Modeling such processes with M3D-$C^1$ is left for future work}. 
We can only remark now that in simulations with two $\iota=1$ surfaces like the one presented here, the inner crash is always immediately followed by an outer crash, which feature does not appear consistent with the ``type-B'' crashes observed.

\section{Summary and discussion}\label{summary}

In this paper, we present nonlinear {single-fluid} MHD simulations of sawtooth-like crashes in the W7-X stellarator.
The simulations are initialized from equilibria with near-axis ECCD given by ray-tracing modeling. The two $\iota=1$ resonances in the rotational transform profile give rise to two consecutive $(1,1)$ internal kink modes in the simulations. 
Reconnection at the inner resonance causes a small-amplitude crash first, which may correspond to the sawtooth precursors observed in the experiments. 
A bigger crash at the outer resonance then flattens the core temperature profile, which shows semi-quantitative agreements with experimental measurements on {certain} metrics such as the crash amplitude and the inversion radius of the temperature change. 
These results suggest that the mechanism of the medium, ``type-A'' sawtooth-like crashes seen in W7-X is likely {reconnection driven by $(1,1)$ internal kink modes}, consistent with \cite{Aleynikova2021}, which provides direct comparison between the experimental observations and the Kadomtsev model.

{To some extent}, this work also validates the newly developed stellarator modeling capability of M3D-$C^1$ \cite{Zhou2021}, and {calibrates} its readiness to undertake meaningful physical studies. {On the one hand, further improvements can be made for more accurate modeling, such as better pre-conditioners for better numerical stability at lower dissipation or more effective inclusion of two-fluid effects. 
On the other hand,} this capability, along with similar developments in JOREK \cite{Nikulsin2019,Nikulsin2021,Nikulsin2022} and NIMROD \cite{Sovinec2021}, enables studies of nonlinear, transport-timescale MHD physics in complex stellarators, {while previous simulations have either shorter timescales \cite{Suzuki2021,Strauss2004,Sato2017} or simpler geometries \cite{Roberds2016,Schlutt2012,Schlutt2013}}. 
One notable opportunity is to investigate the nonlinear stability of stellarator plasmas. 
Stellarator designs are usually constrained by linear MHD stability, but
the plasmas are often experimentally found to be more robust than what linear theory predicts \cite{Weller2006}. Utilizing this feature could expand operation windows for present stellarators and improve designs for future ones.

\section*{Supplementary Material}
{See the supplementary material for a video showing the entire evolution of the magnetic field configuration, which includes the selected snapshots in Figures \ref{inner}, \ref{outer}, and \ref{helical}.}

\acknowledgments
We thank P.~Helander, S.~C.~Jardin, C.~Liu, and A.~M.~Wright for helpful discussions. 
YZ was sponsored by Shanghai Pujiang Program under Grant No.~21PJ1408600 and the Fundamental Research Funds for the Central Universities.
NMF was supported by the U.S. Department of Energy under contract number DE-AC02-09CH11466. 
The United States Government retains a non-exclusive, paid-up, irrevocable, world-wide license to publish or reproduce the published form of this manuscript, or allow others to do so, for United States Government purposes.

\end{document}